\documentclass[Letter]{aa}  

\usepackage{graphicx} 
\usepackage{xcolor}
\usepackage[normalem]{ulem}
\usepackage{hyperref}
\hypersetup{colorlinks,linkcolor={blue},citecolor={Green},urlcolor={blue}}
\usepackage{txfonts}
\usepackage[dvipsnames]{xcolor}
\usepackage{pdflscape}
\usepackage{multirow}
\usepackage{arydshln}
\usepackage{caption}
\usepackage{subcaption}
\usepackage{ulem}

\newcommand{\teff}{T$_{\mathrm{eff}}$}
\newcommand{\msun}{M$_\odot$}
\newcommand{\logg}{log $g$}
\newcommand{\vmic}{$v_{\textrm{mic}}$}

\newcommand{\logeps}{log $\epsilon$}

\newcommand{\sigXoFe}{$\sigma_{\text{[X/Fe]}}$}
\newcommand{\BHs}{\ensuremath{\mathrm{BH3}^{\star}}}

\begin{document} 

    \titlerunning{the {\it Gaia} BH3 system}
    \authorrunning{}
    \title{The chemical fingerprint of the \textbf{\textit{Gaia}} BH3 system}
    \subtitle{Evidence for early cluster enrichment from the analysis of 51 elements \thanks{Based on observations collected at the European Southern Observatory under ESO programme DDT 113.27RA and data obtained from the ESO Science Archive Facility (program ID 112.25ZW).}}

    \author{Gr\'egory Vanden Broeck \inst{1,2} \corrauth{gregory.vanden.broeck@ulb.be} \and
    Thibault Merle \inst{1,2,3} \email{thibault.merle@ulb.be} \and
    Nhat Tan Mai \inst{1, 2, 4} \email{nhat.tan.mai@ulb.be}\and 
    Sophie Van Eck \inst{1,2} \email{sophie.van.eck@ulb.be}\and 
    St\'ephane Goriely \inst{1,2} \email{stephane.goriely@ulb.be}\and 
    Lionel Siess \inst{1,2} \email{lionel.siess@ulb.be}\and
    Alain Jorissen \inst{1,2} \email{alain.jorissen@ulb.be}\and
    Do Thi Hoai \email{dthoai@vnsc.org.vn}\inst{4}}
    
    \institute{Institute of Astronomy and Astrophysics (IAA), Université libre de Bruxelles (ULB), CP 226, Boulevard du Triomphe, 1050 Brussels, Belgium 
    \and {BLU-ULB, Brussels Laboratory of the Universe, blu.ulb.be} 
    \and {Royal Observatory of Belgium, Avenue Circulaire 3, 1180 Brussels, Belgium}
    \and {Vietnam National Space Center (VNSC), Vietnam Academy of Science and Technology (VAST), 18 Hoang Quoc Viet, Nghia Do, Ha Noi, Vietnam}}

    \date{Received; accepted}
     
    \abstract
    {The {\it Gaia} BH3 system hosts the most massive known stellar-origin black hole and a low-mass metal-poor companion whose chemical composition may constrain early explosive nucleosynthesis processes.}
    {We investigate the chemical abundances of the companion in order to constrain the formation of this remarkable system.}
    {We perform a detailed analysis of high-resolution ESO-UVES spectra of the companion. 51 elements from lithium to uranium were investigated through spectral synthesis, including 15 treated in NLTE. We compare the resulting pattern to r-process enriched stars, to nucleosynthesis models and to stars of the ED-2 stream, from which BH3 is thought to originate.}
    {The abundance pattern of the BH3 companion is consistent with that of r-I stars and is well reproduced by a combination of core-collapse supernova yields and an r-process component. 
    The chemical patterns of four ED-2 stars closely match that of the companion, particularly after accounting for different levels of mixing of the enriched material with the ambient gas.}
    {The present analysis provides the most detailed chemical characterisation of a metal-poor star associated with a stellar-mass black hole. The chemical similarity with ED-2 stars  argue against local pollution across the binary system. The abundances instead reflect early spatially inhomogeneous  enrichment of the progenitor cluster.}
  
    
    \maketitle

\nolinenumbers

\section{Introduction}\label{sec:Introduction}

     The {\it Gaia} satellite has revealed binary systems composed of a stellar-mass black hole and an unevolved companion, among which {\it Gaia} BH3 is particularly remarkable \citep{Gaia2024}. It hosts the most massive known stellar-origin black hole ($32.70 \pm 0.82$ \msun), in an eccentric orbit ($P=11.6$ yr, $e=0.73$), and a visible companion, {\it Gaia} DR3 {\texttt source\_id} 4318\-4650\-6642\-052\-8000, which is a low-mass ($0.76 \pm 0.05$ \msun), metal-poor ([Fe/H]\footnote{With the notation  [X/Y]$\equiv\log (N_X/N_Y)_* - \log  (N_X/N_Y)_\odot$ and $\log\epsilon\equiv\log (N_X/N_H)+12$   where $N_X$ is the number density of element $X$.} = $-2.56\pm 0.11$) G-type companion  (hereafter \BHs). Located at $\sim 590$ pc in the Galactic halo, it is dormant, with only upper limits on its emission in the radio \citep{Sjouwerman-2024}, infrared \citep{Kervella-2025}, UV and X-ray bands \citep{Cappelluti-2024, Gilfanov-2024,Sbarufatti-2025}. Its wide orbit excludes Roche-lobe overflow, favouring inefficient accretion processes such as interstellar medium accretion or wind-driven flows \citep{Cappelluti-2024}, with negligible impact on the black hole mass.
     
     The formation of BH3 remains debated. While strong Wolf–Rayet winds hinder the formation of such massive black holes at solar metallicity, massive stars with a metallicity as low as the one of \BHs\ may give rise to $\gtrsim 30$ \msun\ black holes through direct collapse \citep{Merritt-2026}. The orbit of \BHs\ is consistent with the ED-2 stream, likely originating from a disrupted cluster of mass $2 \times 10^3$–$5.2 \times 10^4$ \msun\ \citep{Balbinot2024}, motivating scenarios involving dynamical interactions or mergers \citep{Balbinot2024, Marin-Pina-2024}.
     
     The chemical composition of \BHs\ provides key constraints on its origin. It has been studied by \citet{Gaia2024} and \citet[][hereafter H25]{Hackshaw2025}.  In this Letter, we perform a comprehensive reanalysis of the abundance pattern using high-resolution UVES spectra, deriving stellar parameters in non-Local Thermodynamic Equilibrium (NLTE) and applying NLTE corrections to abundances when available. We compare the resulting abundances with nucleosynthesis predictions and the assumed-universal solar r-process pattern and discuss possible formation scenarios.

\section{Data}\label{sec:data}
    
    The spectra were obtained with UVES (proposal DDT 113.27RA). Observations were carried out on May 20 to 22, 2024 using dichroic \#1 and dichroic \#2 with the standard UVES settings 346 + 580 and 437 + 860, respectively, covering the optical range from 305 to 1040 nm. A slit width of $0.4-0.5''$ was adopted, yielding a spectral resolution of $R = 65\,000$ for the 346 and 437 settings, and $R = 87\,000$ for the 580 and 860 settings. By co-adding three 2700s exposures in each setting, typical signal-to-noise ratio (S/N) values of 66 and 200 are achieved at wavelengths 3600\,\AA\ and 5300\,\AA, respectively. The average radial velocity is $-379.67\pm 0.29$ km~s$^{-1}$, corresponding to a mean JD of 2\,460\,452.127.

\section{Stellar parameter determination}
\label{sec:stellar_param_determination}
    \begin{figure}
        \centering
        \includegraphics[width=\columnwidth, trim=5 7 5 7, clip]{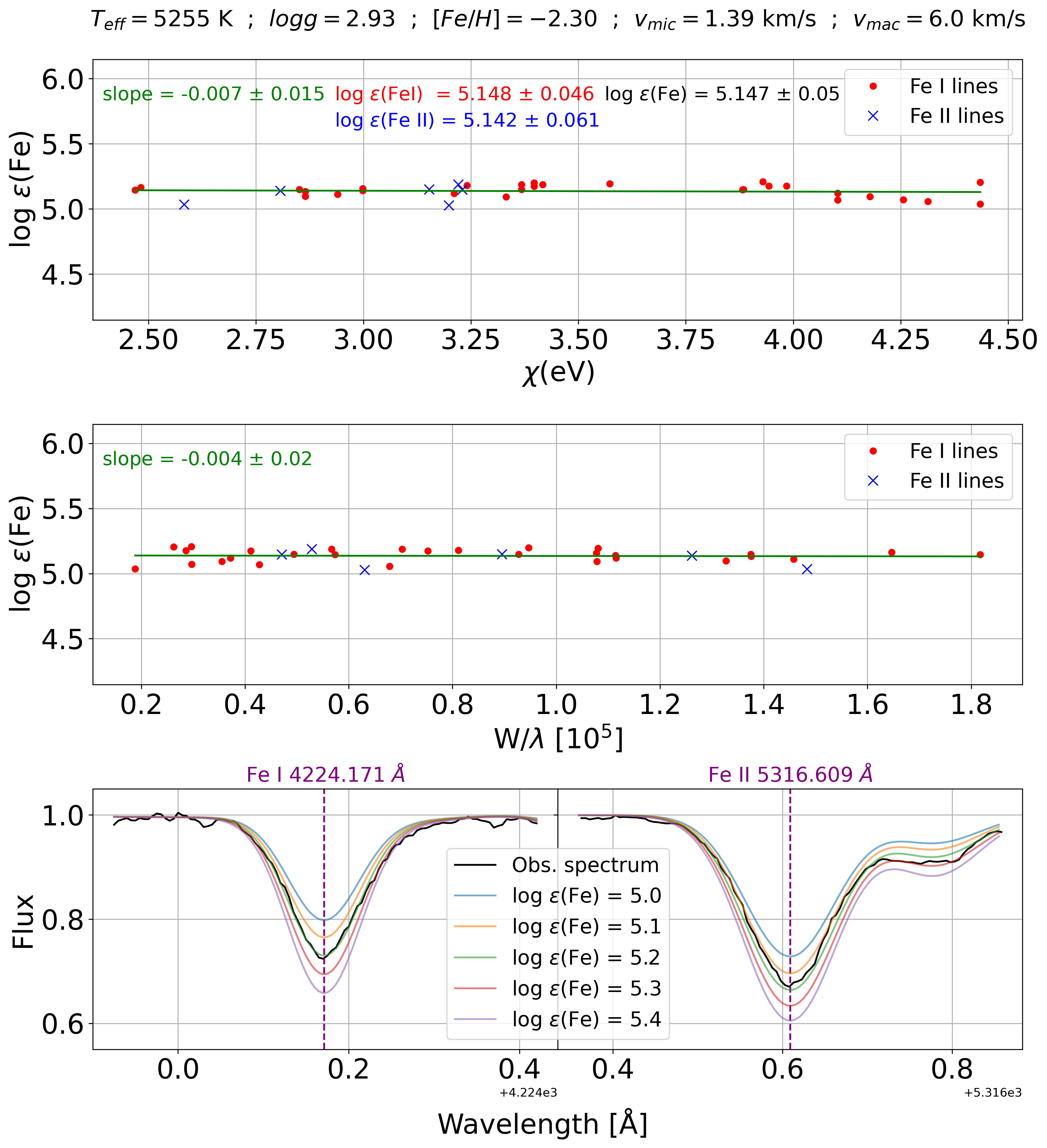}
        \caption{\tiny Top panel:  NLTE \ion{Fe}{i} and \ion{Fe}{ii} abundances as a function of excitation potential. Middle panel: Same as above, plotted as a function of reduced equivalent width. Bottom panels: example fits of \ion{Fe}{i} and \ion{Fe}{ii} lines for which $\chi^2$ minimization yields abundances of $\log\epsilon({\ion{Fe}{i}}) = 5.19$ and  $\log\epsilon({\ion{Fe}{ii}}) = 5.15$.} 
        \label{fig:stellar_param_determination}
    \end{figure}

\begin{figure}[!t]
        \centering
        \includegraphics[width=\columnwidth, trim=10 7 7 7, clip]{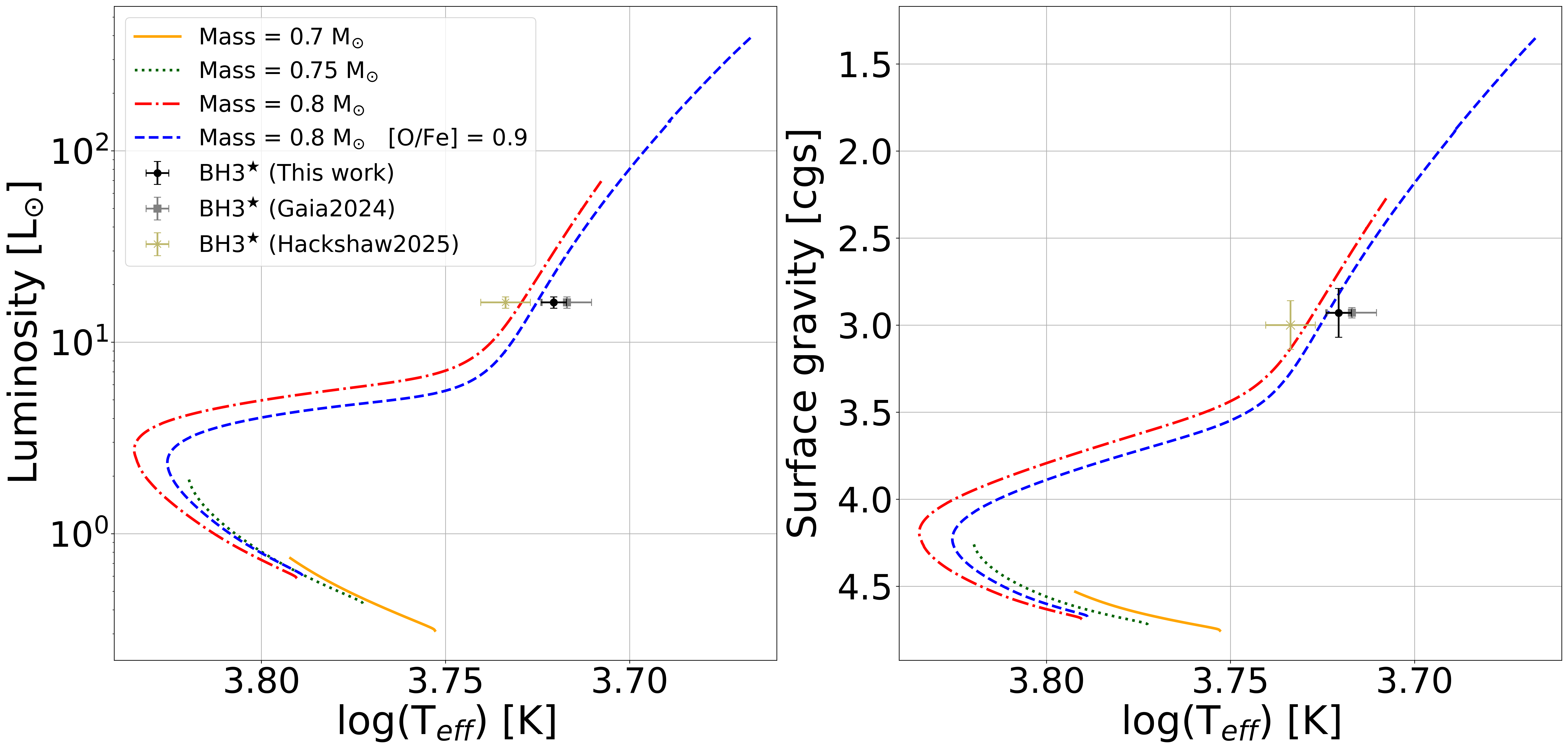}
        \caption{\tiny Position of \BHs\ in the HR (\teff, L) and Kiel  (\teff, \logg) diagrams along with STAREVOL evolutionary tracks computed with a metallicity [Fe/H] $=-2.3$, for various initial masses and either solar or enriched [O/Fe] values, as indicated.}
        \label{fig:Kiel_diagram}
    \end{figure}

    Synthetic spectra were computed using hydrostatic MARCS model atmospheres \citep{Gustafsson-2008} and the LTE/NLTE (Appendix \ref{appendix:nlte_corrections}) Turbospectrum code \citep{Gerber-2023, Turbospectrum2020} to derive Fe abundances from 35 carefully selected \ion{Fe}{i} and \ion{Fe}{ii} lines (Table~\ref{appendix:atomic_line_info}), and analysed as functions of excitation potential and reduced equivalent width (Fig.~\ref{fig:stellar_param_determination}). Following \cite{Bergemann-2012b}, only lines with excitation potentials above 2 eV were used to minimise 3D effects. Stellar parameters and their errors were determined as described in Appendix~\ref{Sect: Appendix-parameters}. The final parameters are listed in Table~\ref{tab:stellar_parameters}. Our \teff\ and $\log g$ agree with \cite{Gaia2024}, while [Fe/H] and \vmic\ are closer to H25. This reflects methodological differences:  H25  used photometric calibrations \citep{Mucciarelli-2021, Casagrande-2014} combined with LTE spectroscopy (\textsc{BACCHUS}; \citealt{Masseron-2016}), while we adopt a fully spectroscopic NLTE approach based on excitation and ionisation equilibrium. Classical 1D LTE methods are known to be biassed at low metallicity due to NLTE effects on \ion{Fe}{i} lines \citep{Casagrande-2010, Frebel-2013}. In contrast, NLTE analyses provide results consistent with astrometric gravities \citep{Ruchti-2013} and less dependent on photometric calibrations. Parameter uncertainties are significantly reduced (excluding comparison with the unrealistically small 0.003 dex error in $\log g$ from \citealt{Gaia2024}), especially for metallicity. As shown in Fig.~\ref{fig:Kiel_diagram}, our parameters are consistent with a 0.8\msun{} STAREVOL track \citep{Siess-2006} with an enhanced [O/Fe] = 0.9, consistent with \BHs\ (Table~\ref{tab:chemical_abundances}).

    \begin{table}
        \centering
        \caption{\BHs\ stellar parameters}
        \begin{tabular}{lccc}
                \hline
                \hline \\ [-0.25cm]
                Parameter & This work & Gaia2024 & H25 \\
                 & (NLTE) & (LTE) & (LTE) \\
                \hline \\ [-0.25cm]
                \teff\ [K] & $5255 \pm 65$ & $5212 \pm 80$ & $5416 \pm 84$ \\
                \logg & $2.93 \pm 0.21$ & $2.929 \pm 0.003$ & $3.00 \pm 0.14$ \\
                $\mathrm{[Fe/H]}$ & $-2.30 \pm 0.12$ & $-2.56 \pm 0.11$ & $-2.27 \pm 0.24$  \\
                \vmic\ [km/s] & $1.39 \pm 0.07$ & $1.19$ & $1.54 \pm 0.11$ \\
                \hline
            \end{tabular}
        \tablefoot{Effective temperature (\teff), surface gravity (\logg), metallicity ([Fe/H]), and microturbulent velocity (\vmic), are listed as derived in this work and as reported by \cite{Gaia2024} and H25.}
                \label{tab:stellar_parameters}
    \end{table}

\section{Chemical abundances derivation}\label{sec:chemical_abund_determination}

    \begin{figure*}
        \centering
        \includegraphics[width=0.9\linewidth]{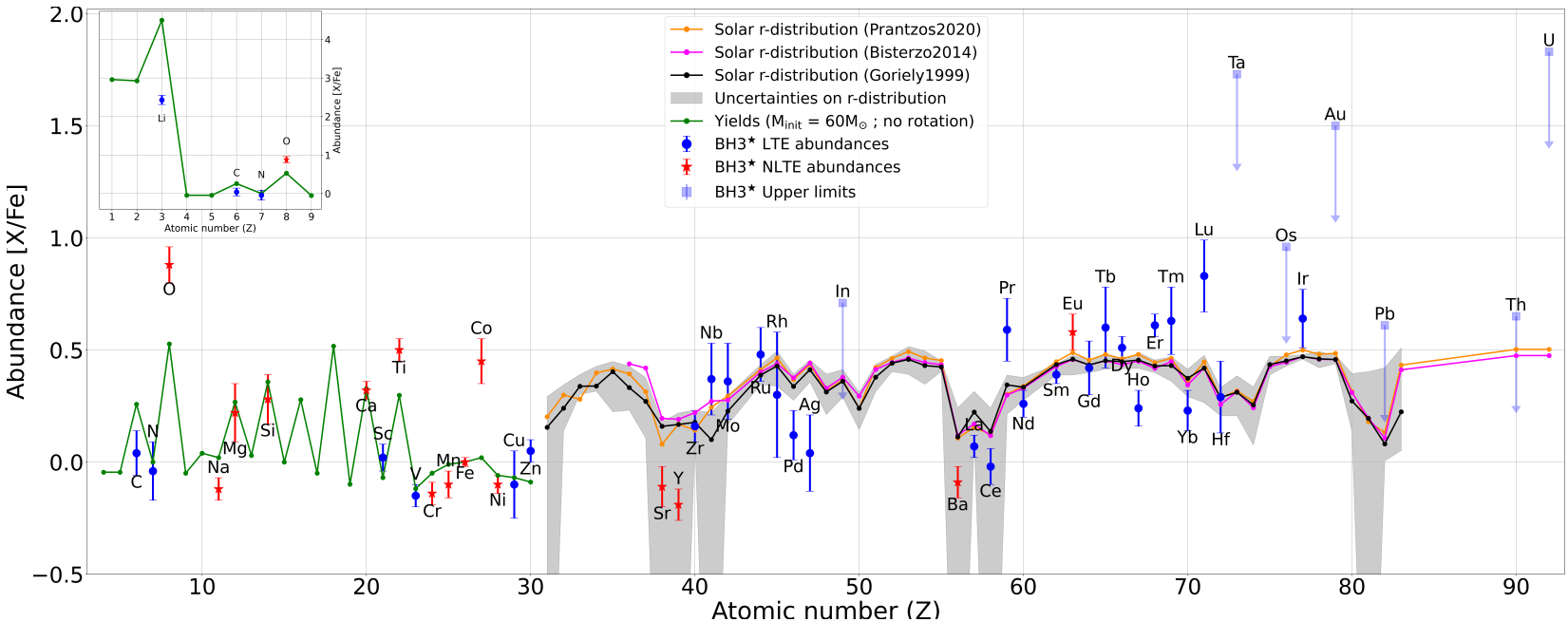}
        \caption{\tiny Chemical abundances of \BHs\ compared (i) for $Z\le30$, with yields of a [Fe/H] $=-3$, 60~\msun\, non-rotating ccSN \citep{Limongi2018} and (ii) for $Z>30$, with solar r-process distributions determined by \cite{Goriely1999}, \cite{Bisterzo2014} and \cite{Prantzos2020}. The theoretical profiles are scaled to minimize the $\chi^2$ between measured and predicted abundances of heavy (Z $\geq 38$) elements. The grey shaded region represents the uncertainties associated with the predictions of \cite{Goriely1999}. For clarity, the measured and predicted Li abundance are shown in the inset as the [Li/Fe] ratio is higher than for the other elements.}
        \label{fig:AbProfile_final_comparison}
    \end{figure*}

    51 chemical elements from Li to U were investigated, with only upper limits determined  for In, Ta, Os, Au, Pb, Th and U. Solar abundances are from \cite{Grevesse-2007}. The abundances were derived by minimising the $\chi^2$ between the observed and synthetic profiles computed with Turbospectrum in NLTE when a model atom was available (Appendix \ref{appendix:nlte_corrections}), using a polynomial fit to at least six bracketing abundance values.
    
    The abundances derived here agree with those of H25 within the uncertainties, except for O (3 NLTE vs 1 LTE line), V (10 vs 1), Mn (15 NLTE vs 1 LTE), Co (6 NLTE vs 1 LTE), Ni (72 NLTE vs 9 LTE), La (9 vs 3), Pr (6 vs 1) and Sm (9 vs 1). For Cu, the uncertainty ranges overlap only marginally: the \ion{Cu}{i} 5153.227~\AA\ line used by H25 is not detected in our spectra, so we instead fitted the wings of the strong \ion{Cu}{i} 3273.954~\AA\ line. Our analysis also extends H25 by including the following additional elements: Nb, Mo, Ru, Rh, Pd, Ag, In, Gd, Tb, Ho, Er, Tm, Yb, Lu, Hf, Ta, Os, Ir, Au, Pb, and U. For each element, uncertainties in the stellar parameters (Table~\ref{tab:stellar_parameters}) were propagated to the abundances by varying each parameter within its uncertainty range, following \cite{Johnson-2002}, including the effect of [Fe/H] and neglecting the covariance terms, thus providing a conservative estimate. Table~\ref{tab:chemical_abundances} lists the resulting abundances and uncertainties.

\section{\BHs\ abundance profile and comparison with nucleosynthesis models}\label{sec:Comparison_models}

    The absence of overabundance in s-process elements (Sr, Y, Ba, La, Ce) points toward an r-process origin, as expected for a low-metallicity, old object such as \BHs\ (Sect.~\ref{sec:ED2}), where enrichment by previous generations of low or intermediate-mass stars has not yet occurred.  From our NLTE Ba and Eu abundances, we confirm the classification of \BHs\ as an r-I star ($0.3 \le {\rm [Eu/Fe]} \le 1$, ${\rm [Ba/Eu]} < 0$; \citealt{Beers2005, Masseron2010}). To confirm this classification, we compare the abundance pattern of \BHs\ with average r-I and r-II patterns and with the weak r-process stars HD~88609 \citep{Honda_2007} and HD~122563 \citep{Honda_2006} (Figs.~\ref{fig:AbProfile_vs_rIIstars} and \ref{fig:AbProfile_vs_Honda}). The abundance pattern of \BHs\ closely matches that of r-I stars. The main discrepancies (e.g. O, Ca, Ti, Co) involve elements derived in NLTE here, whereas literature abundances are mostly based on LTE analyses. Concerning cobalt, the discrepancy between model predictions and measured Co abundances in metal-poor stars is a long-standing and still unsolved issue \citep{Bergemann-2010}. In Fig.~\ref{fig:AbProfile_final_comparison}, we compare the abundance pattern of \BHs\ with solar r-process distributions from \citet{Goriely1999, Bisterzo2014, Prantzos2020}. The good agreements between the abundances of r-I stars and those of \BHs, on the one hand, and between r-I stars and the solar r-process distribution on the other hand \citep{Roederer-2010, Cowan-2021}, also imply that  \BHs\ abundance pattern is compatible with the solar r-process distribution. Notable offsets are found for Sr$^{\rm NLTE}$, Y$^{\rm NLTE}$, Ba$^{\rm NLTE}$, and Ce$^{\rm LTE}$, but remain compatible given the large uncertainties affecting the solar r-contribution to these elements \citep{Goriely1999}. Additional discrepancies occur for Pd (1), Ag (1), Pr (1), Ho (3), Er (5), and Lu (1) (number of lines in parentheses), whose abundances lie outside the  combined measured and theoretical uncertainties. These may reflect limited line statistics, NLTE effects, or shortcomings in our present nucleosynthesis understandings. We also compare the abundances with core-collapse supernova (ccSN)  yields at [Fe/H]$=-3$ from \cite{Limongi2018}. The best agreement is obtained for non-rotating models with progenitor masses $40 \leq M \leq 120$~\msun\ and no dilution of the ejecta into a solar composition scaled to \BHs\ metallicity (Fig.~\ref{fig:AbProfile_final_comparison} and \ref{fig:AbProfile_vs_Yields}). The abundances of \BHs\ are thus compatible with an enrichment by probably only a few ccSNe events as far as light elements are concerned and with an r-process pollution for the elements heavier than iron.

\section{Comparison with four other ED-2 stars }\label{sec:ED2}
    \begin{figure}
        \centering
        \includegraphics[width=0.9\columnwidth, trim=7 8 7 5, clip]{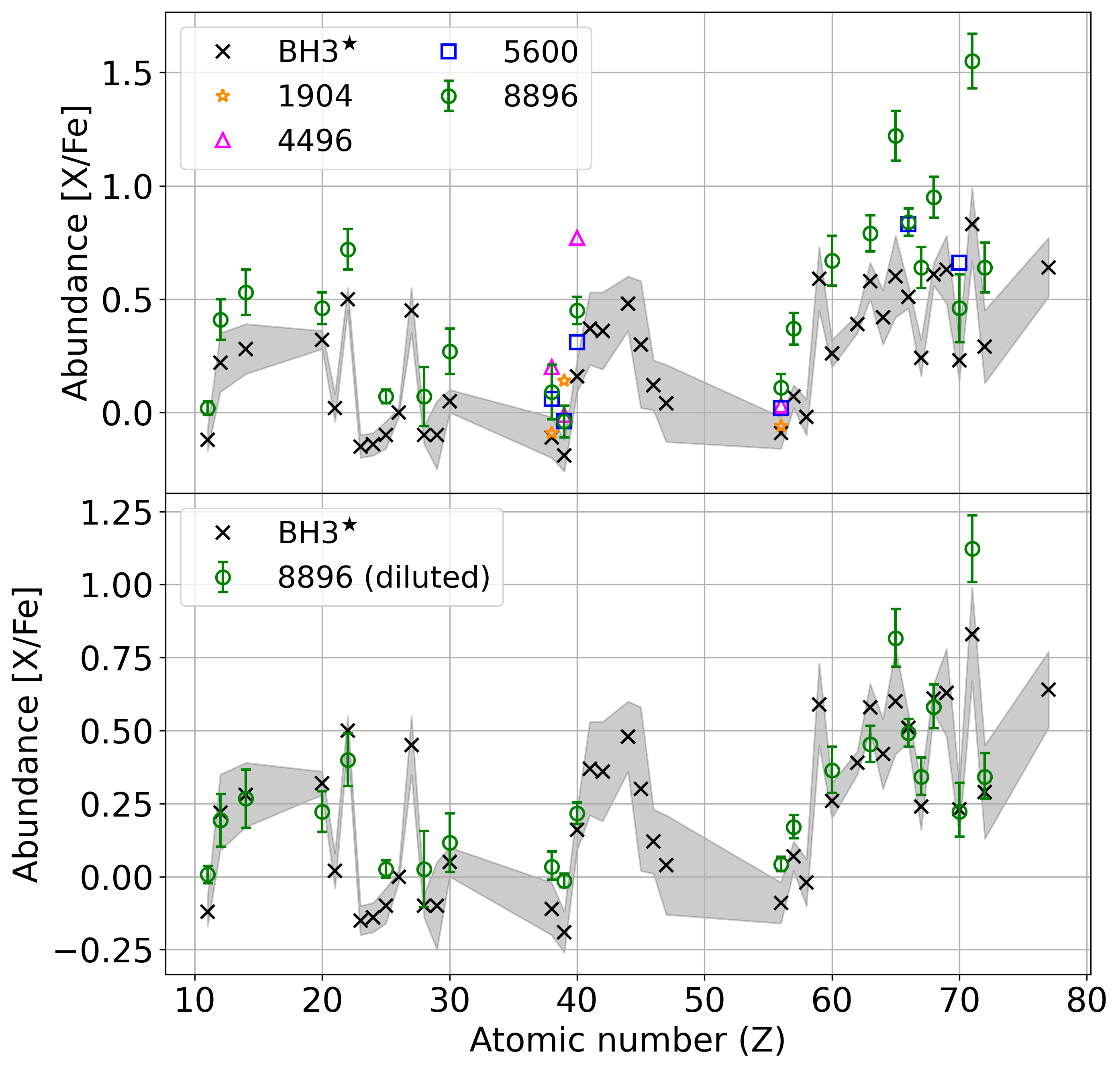}
        \caption{\tiny Upper panel: Abundance profile of \BHs\ compared to abundances of four ED-2 stars. The stars {\it Gaia} DR3 
        \texttt{source\_id} 4245\-5224\-6855\-409\-1904, 4479\-2263\-1075\-831\-4496, 6632\-3350\-6023\-108\-8896 and 6746\-1145\-8505\-626\-5600 are identified by their last four digits. The uncertainties on the \BHs\ abundances are shown by the grey shaded region.
         Lower panel: Comparison between \BHs\ and star 8896, after diluting the abundances of star 8896 by a factor of 0.7 into a solar composition scaled to the metallicity of \BHs.}
        \label{fig:AbProfile_vs_ED2stars}
    \end{figure}
    
    Unlike BH1 and BH2, BH3 is associated with the Galactic halo. Its companion \BHs\ is a high–proper-motion star on a strongly retrograde orbit, whose kinematics and low metallicity suggest a globular cluster origin \citep{Balbinot2024}. It is linked to the ED-2 stream, likely the remnant of a disrupted cluster among the oldest in the Galaxy, based on comparison with M92 ($13.80\pm0.75$ Gyr). 
    
    Using archival ESO-UVES spectra, we derived abundances for four ED-2 stars (Appendix~\ref{Sect: Appendix-ED2}). Figure~\ref{fig:AbProfile_vs_ED2stars} shows that all four stars closely match the abundance pattern of \BHs. The similar masses of \BHs\ ($0.76 \pm 0.05$~\msun) and {\it Gaia}~DR3 \texttt{source\_id}  6632\-3350\-6023\-108\-8896 (hereafter 8896) (from \cite{Balbinot2024} \teff$ = 5620$~K, \logg $= 3.60$, placing it on the [O/Fe]-enhanced $0.8$~\msun\ track in Fig.~\ref{fig:Kiel_diagram}) contrast with their differing levels of overabundances. With 14 heavy elements measured, star 8896 agrees remarkably well with \BHs\ when a dilution factor of 0.7 is applied (meaning 30\% star-8896 material mixed with 70\% solar-scaled material). The fact that the abundance profiles (but not the overabundance levels), are similar, suggests that the enrichment within the cluster later disrupted as ED-2 was chemically, but not spatially, homogeneous.

\section{Discussion and conclusion}\label{sec:discussion}
    Several scenarios may explain the abundance peculiarities of \BHs. Given ccSN ejecta velocities of $10^3$–$10^4$ km/s, Bondi–Hoyle accretion onto \BHs\ is negligible \citep{Liu-2015}, rendering significant pollution by the BH3 progenitor unlikely. The similar abundance patterns observed among ED-2 stars instead favour a common enrichment origin. We refer to the disrupted progenitor of the ED-2 stream as the “ED-2 cluster”. Given the current observational constraints, it remains unclear whether (i) a ccSN both formed BH3 and enriched the gas from which the ED-2 cluster originated, or (ii) a ccSN enriched this gas while BH3 formed later, or (iii) the ED-2 cluster formed from pristine gas and was subsequently enriched by a ccSN that produced BH3. The number of ccSNs required must be limited given the age of ED-2 \citep{Balbinot2024}. It is unknown whether \BHs\ formed as a companion to the BH3 progenitor, consistent with the simulations of \cite{Iorio-2024}, or was later dynamically captured, a scenario favoured by \cite{El-Badry-2024}. Indeed, the high eccentricity of \BHs\ ($e=0.73$) does not exclude the capture scenario by BH3, as dynamical interactions in dense systems can form eccentric binaries \citep{Portegies-2008}. More complex scenarios could be possible in particular if BH3 is itself a tight binary composed of two black holes.
    
    In conclusion, the present analysis, based on 51 elements, provides the most detailed chemical characterization of a metal-poor star associated with a stellar-mass black hole. The abundance pattern confirms its r-I nature and is consistent with enrichment from ccSNs for light elements and an r-process contribution for the elements heavier than iron. The agreement with ED-2 stars reflect early cluster-wide, spatially inhomogeneous enrichment rather than a direct contamination from the black hole progenitor, linking nucleosynthesis, cluster evolution, and black hole formation.

\section*{Data availability}
Table~\ref{tab:atomic_line_info} is only available in electronic form at the CDS via anonymous ftp to cdsarc.u-strasbg.fr (130.79.128.5) or via \url{http://cdsweb.u-strasbg.fr/cgi-bin/qcat?J/A+A/}

\begin{acknowledgements}
GVB acknowledges a ULB postdoctoral fellowship; T.M., BELSPO FED-tWIN support (Prf-2020-033\_BISTRO); MT, an ARES PhD fellowship; SVE, Fondation ULB; and SG and LS, F.R.S.-FNRS support. This work was funded by F.R.S.-FNRS and FWO through EOS project O000422F and used the SIMBAD database operated by CDS, Strasbourg, France. We thank the referee for constructive comments that improved the paper.
\end{acknowledgements}
\bibliographystyle{aa}
\bibliography{library}

\clearpage 
\newpage  
\nolinenumbers

\begin{appendix}
\nolinenumbers

\section{NLTE corrections}\label{appendix:nlte_corrections}

    The NLTE model atoms adopted in this work are described in the following studies: 
    H \citep{Mashonkina-2008}, 
    O \citep{Bergemann-2021}, 
    Na \citep{Larsen-2022}, 
    Mg \citep{Bergemann-2017}, 
    Si \citep{Bergemann-2013, Magg-2022}, 
    Ca \citep{Mashonkina-2017, Semenova-2020}, 
    Ti \citep{Bergemann-2011}, 
    Mn \citep{Bergemann-2019}, 
    Fe \citep{Bergemann-2012b, Semenova-2020}, 
    Co \citep{Bergemann-2010, Yakovleva-2020}, 
    Ni \citep{Bergemann-2021, Voronov-2022}, 
    Sr \citep{Bergemann-2012a}, 
    Ba \citep{Gallagher-2020}, 
    Y \citep{Storm-2023}, 
    Eu \citep{Storm-2024}

\section{Linelist}\label{appendix:atomic_line_info}

    \begin{table}[ht]
        \caption{Linelist used for the analysis of \BHs\ and for the four ED-2 stars.}
        \centering
        \hspace{-1cm}
        \begin{minipage}{0.25\textwidth}
            \centering
            \begin{tabular}{@{\hspace{15pt}}c@{\hspace{15pt}}c@{\hspace{15pt}}c@{\hspace{15pt}}}
                \hline\hline \\[-0.3cm]
                $\lambda$ (\AA) & $\chi$ (eV) & log $gf$ \\[0.1cm]
                \hline \\[-0.2cm]
                \multicolumn{3}{c}{\textbf{\ion{Li}{i}}} \\[0.058cm]
                6707.764 & 0.000 & $-0.002$ \\ [0.058cm]
                \hline \\[-0.25cm]
                \multicolumn{3}{c}{\textbf{\ion{O}{i}}} \\[0.058cm]
                7771.940 & 9.146 & 0.349 \\ [0.058cm]
                ...      &  ...  &  ...  \\[0.058cm]
                \hline
            \end{tabular}
        \end{minipage}
         \tablefoot{The complete table is available in electronic form at the CDS.}
    \label{tab:atomic_line_info}
    \end{table}

\section{Stellar parameter determination}
\label{Sect: Appendix-parameters}

    Varying the stellar parameters affects five independent diagnostics: the slopes of the Fe abundance trends with excitation potential and reduced equivalent width, the agreement between the mean Fe I and Fe II abundances, the consistency between the input metallicity and the metallicity inferred from the Fe abundance, and the dispersion of the Fe abundances. We define a dimensionless merit function by normalizing each diagnostic to its estimated uncertainty and summing the resulting terms with equal weight. The optimal stellar parameters are taken to be those that minimize this merit function, thereby simultaneously enforcing excitation equilibrium, ionization equilibrium, metallicity consistency, and minimal abundance scatter.
    
    The stellar parameters were first explored on a grid spanning 4700K $\le$ \teff\ $\le$ 5700K, 2.5 $\le$ \logg\ $\le$ 3.4, 0.7 $\le v_{\rm mic} \le$ 1.6 km~s$^{-1}$, with steps of 100K, 0.1 dex and 0.1 km~s$^{-1}$, respectively. A second iteration was then performed on a refined grid centred on the best-fit values, with parameters listed in Table~\ref{tab:stellar_parameters_tested}.
    
    The uncertainties on the stellar parameters were estimated from the variation of the merit function around its minimum. The uncertainty on the merit function was obtained by propagating the normalized uncertainties of the Fe abundance slopes and the Fe abundance dispersion. The $1\sigma$ uncertainty on each parameter was then defined as the variation required to increase the merit function by this amount relative to its minimum. 

\vspace{2cm}

    \begin{table}[!ht]
        \centering
        \caption{Stellar parameters tested in this work.}
            \begin{tabular}{cccccc}
                \hline
                \hline \\
                [-0.2cm]
                Parameter & \multicolumn{5}{c}{Values} \\
                [0.1cm]
                \hline \\
                [-0.1cm]
                \teff [K] & 5200 & 5250 & 5300 & 5350 & 5400\\ [0.1cm]
                \logg & 2.90 & 3.00 & 3.05 & 3.10 & 3.20 \\ [0.1cm]
                [Fe/H] & $-2.20$ & $-2.25$ & $-2.30$ & $-2.35$ & $-2.40$ \\ [0.1cm]
                \vmic [km/s] & 1.36 & 1.38 & 1.40 & 1.42 & 1.44 \\ [0.2cm]
                \hline
            \end{tabular}
           \label{tab:stellar_parameters_tested}
    \end{table}

\section{Comparison of the r-I abundance profile of \BHs\ to r-process enriched stars}\label{appendix:comp-r}

    \begin{figure}[!ht]
        \centering
        \includegraphics[width=\linewidth]{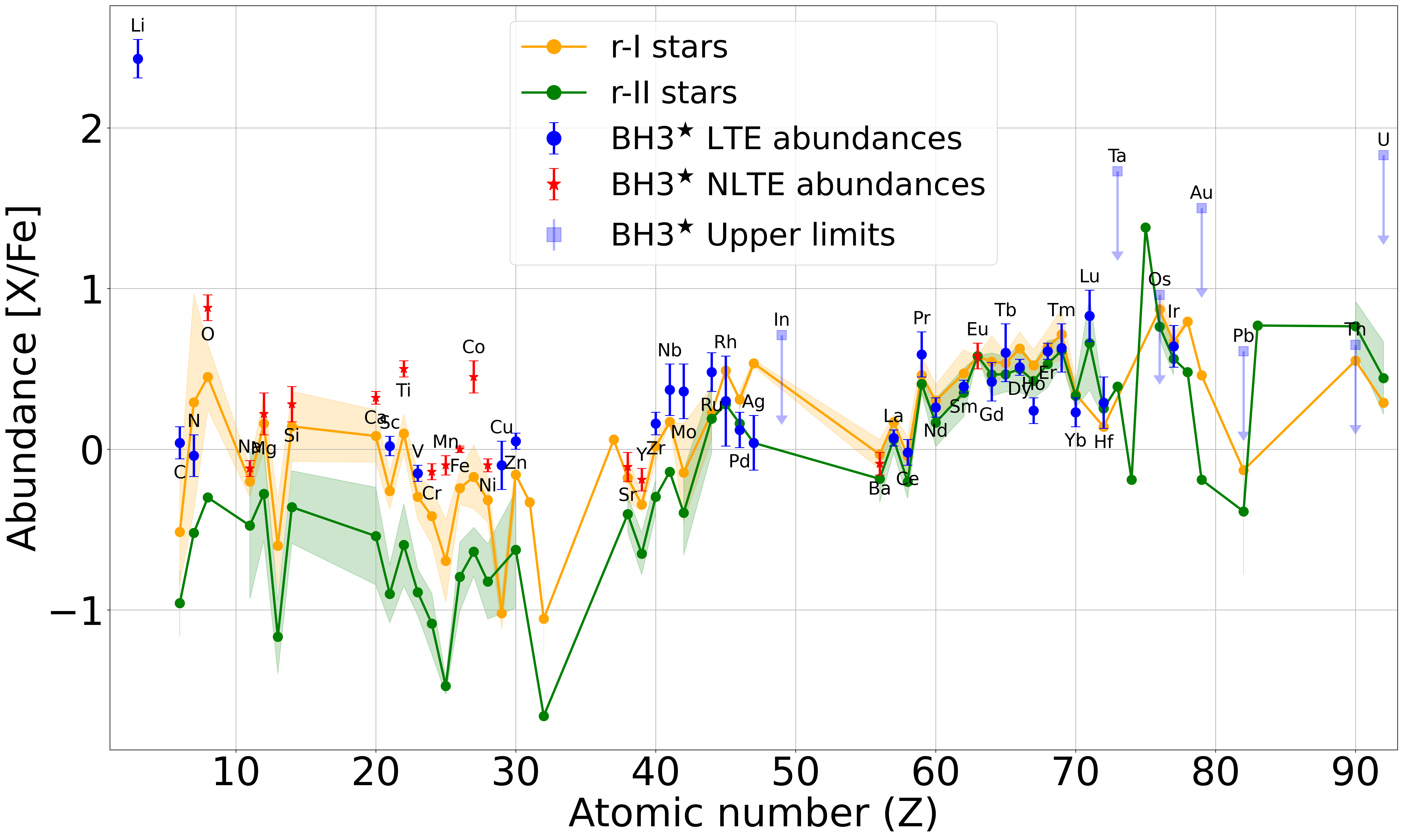}
        \caption{\tiny Abundance pattern of \BHs\ compared with the average abundance patterns of r-I (yellow shaded region) and r-II (green shaded region) stars. The r-I sample is composed of CS 29491–069 \citep{Hayek2009}, HD 115444 \citep{Westin2000}, HD 221170 \citep{Ivans2006}, HE 2252-4225 \citep{Mashonkina2014} and BD~$+17^\circ3248$ \citep{Cowan2002}. The r-II sample is composed of RAVE J203843.2–002333 \citep{Placco2017}, HE 1219–0312 \citep{Hayek2009}, CS 31082-001 \citep{Siqueira-Mello2013}, 2MASS J09544277+5246414 \citep{Holmbeck2018} and LAMOST J110901.22+075441.8 \citep{Li2015}. The r-I and r-II abundance patterns are normalized to Eu.}
        \label{fig:AbProfile_vs_rIIstars}
    \end{figure}
    
    \begin{figure}[!ht]
        \centering
        \includegraphics[width=\linewidth]{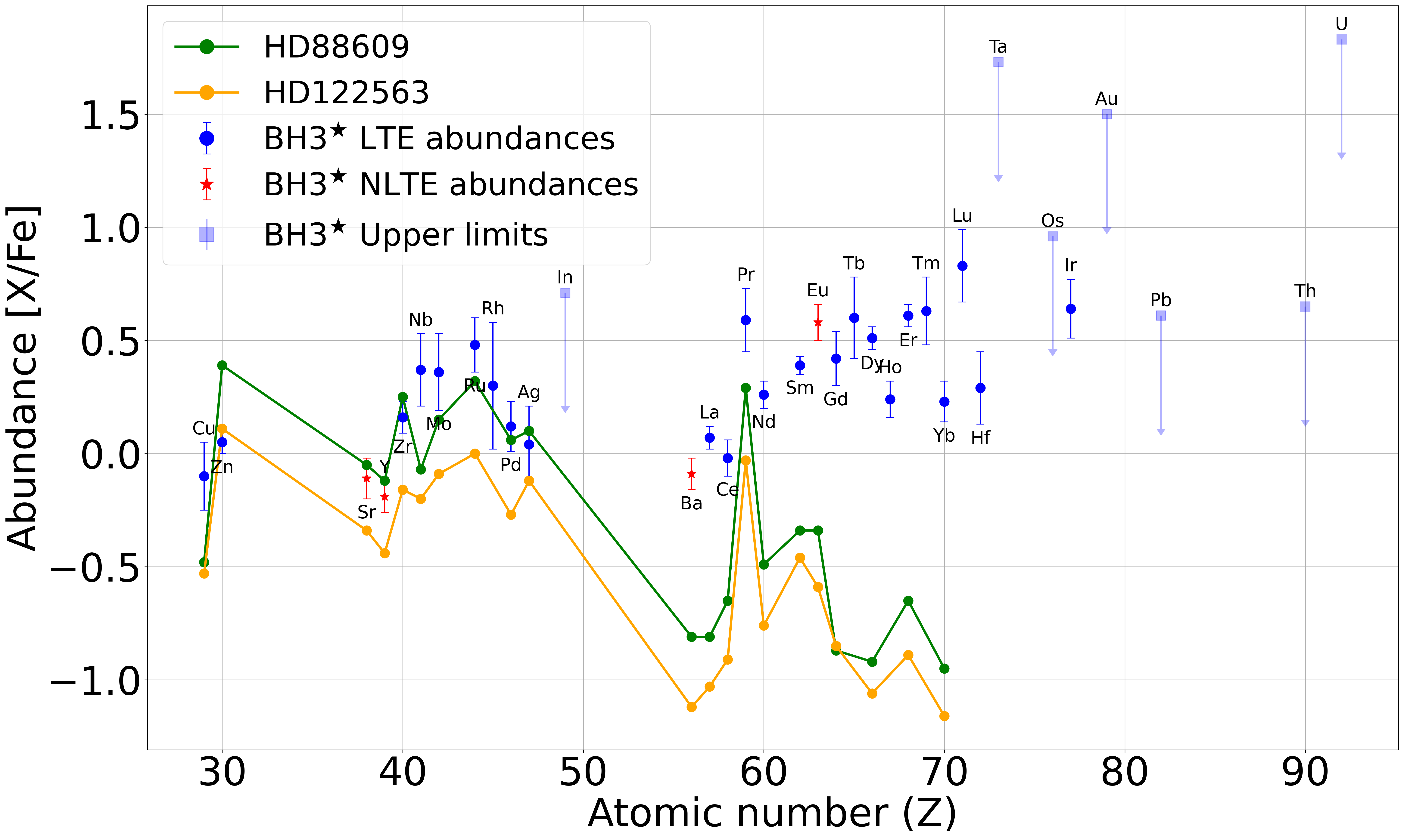}
        \caption{\tiny Abundance profile of \BHs\ compared with two 'weak' r-process stars of \cite{Honda_2006} and \cite{Honda_2007}}.
        \label{fig:AbProfile_vs_Honda}
    \end{figure}

\clearpage 

\onecolumn
\section{\BHs\ chemical abundances}

    \begin{table*}[!ht]
        \centering
        \caption{Chemical abundances of \BHs\ derived in this work.}
        \label{tab:chemical_abundances}
        
        \begin{tabular}{@{}c@{\hspace{9pt}}l@{\hspace{9pt}}c@{\hspace{9pt}}c@{\hspace{9pt}}c@{\hspace{9pt}}c@{\hspace{9pt}}c@{\hspace{9pt}}c@{}}
            \hline\hline
            \rule{0pt}{2.6ex} Z & Elem. & N & \logeps & $\sigma_{\rm red}$ & $\sigma_{\rm tot}$ & [X/Fe] & \sigXoFe \\ [0.1cm]
            \hline \\
            [-0.1cm]
            3  & \ion{Li}{i}                                & 1   & 1.17  & 0.10 & 0.12 & 2.43    & 0.12 \\ [0.053cm]
            6  & \ion{C}{i}                                 & 9   & 6.12  & 0.01 & 0.10 & 0.04    & 0.10 \\ [0.053cm]
            7  & \ion{N}{i}                                 & 1   & 5.43  & 0.10 & 0.13 & $-0.04$ & 0.13 \\ [0.053cm]
            8  & \ion{O}{i} $^{\textsc{NLTE}}$              & 3   & 7.23  & 0.03 & 0.08 & 0.88    & 0.08 \\ [0.053cm]
            11 & \ion{Na}{i} $^{\textsc{NLTE}}$             & 5   & 3.74  & 0.04 & 0.05 & $-0.12$ & 0.05 \\ [0.053cm]
            12 & \ion{Mg}{i} $^{\textsc{NLTE}}$             & 8   & 5.44  & 0.03 & 0.13 & 0.22    & 0.13 \\ [0.053cm]
            14 & \ion{Si}{i} $^{\textsc{NLTE}}$             & 17  & 5.48  & 0.04 & 0.10 & 0.28    & 0.10 \\ [0.053cm]
            14 & \ion{Si}{ii} $^{\textsc{NLTE}}$            & 1   & 5.69  & 0.10 & 0.18 & 0.49    & 0.18 \\ [0.053cm]  
            14 & $\langle$ Si $\rangle$ $^{\textsc{NLTE}}$  & 18  & 5.48  & 0.04 & 0.11 & 0.28    & 0.11 \\ [0.053cm]
            20 & \ion{Ca}{i} $^{\textsc{NLTE}}$             & 49  & 4.31  & 0.01 & 0.03 & 0.31    & 0.04 \\ [0.053cm]  
            20 & \ion{Ca}{ii} $^{\textsc{NLTE}}$            & 2   & 4.44  & 0.01 & 0.07 & 0.44    & 0.07 \\ [0.053cm]  
            20 & $\langle$ Ca $\rangle$ $^{\textsc{NLTE}}$  & 51  & 4.32  & 0.01 & 0.03 & 0.32    & 0.04 \\ [0.053cm]
            21 & \ion{Sc}{ii}                               & 19  & 0.88  & 0.01 & 0.06 & 0.02    & 0.06 \\ [0.053cm]
            22 & \ion{Ti}{i} $^{\textsc{NLTE}}$             & 63  & 3.13  & 0.02 & 0.05 & 0.54    & 0.05 \\ [0.053cm]
            22 & \ion{Ti}{ii} $^{\textsc{NLTE}}$            & 59  & 3.08  & 0.01 & 0.06 & 0.49    & 0.06 \\ [0.053cm]
            22 & $\langle$ Ti $\rangle$ $^{\textsc{NLTE}}$  & 122 & 3.09  & 0.01 & 0.05 & 0.50    & 0.05 \\ [0.053cm]
            23 & \ion{V}{i}                                 & 5   & 1.47  & 0.02 & 0.12 & $-0.22$ & 0.12 \\ [0.053cm]  
            23 & \ion{V}{ii}                                & 5   & 1.59  & 0.01 & 0.06 & $-0.10$ & 0.06 \\ [0.053cm] 
            23 & $\langle$ V $\rangle$                      & 10  & 1.54  & 0.02 & 0.05 & $-0.15$ & 0.05 \\ [0.053cm]
            24 & \ion{Cr}{i} $^{\textsc{NLTE}}$             & 19  & 3.19  & 0.03 & 0.05 & $-0.14$ & 0.05 \\ [0.053cm]
            25 & \ion{Mn}{i} $^{\textsc{NLTE}}$             & 15  & 2.98  & 0.04 & 0.06 & $-0.10$ & 0.06 \\ [0.053cm]
            26 & \ion{Fe}{i} $^{\textsc{NLTE}}$             & 29  & 5.14  & 0.01 & 0.02 & 0.00    & 0.02 \\ [0.053cm]  
            26 & \ion{Fe}{ii} $^{\textsc{NLTE}}$            & 6   & 5.13  & 0.02 & 0.03 & -0.01   & 0.02 \\ [0.053cm]  
            26 & $\langle$ Fe $\rangle$ $^{\textsc{NLTE}}$  & 35  & 5.14  & 0.01 & 0.02 & 0.00    & 0.02 \\ [0.053cm]
            27 & \ion{Co}{i} $^{\textsc{NLTE}}$             & 6   & 3.06  & 0.06 & 0.10 & 0.45    & 0.10 \\ [0.053cm]
            28 & \ion{Ni}{i} $^{\textsc{NLTE}}$             & 72  & 3.82  & 0.01 & 0.04 & $-0.10$ & 0.04 \\ [0.053cm]
            29 & \ion{Cu}{i}                                & 1   & 1.80  & 0.10 & 0.15 & $-0.10$ & 0.15 \\ [0.053cm]
            30 & \ion{Zn}{i}                                & 6   & 2.34  & 0.03 & 0.05 & 0.05    & 0.05 \\ [0.053cm]
            38 & \ion{Sr}{i} $^{\textsc{NLTE}}$             & 1   & 0.51  & 0.10 & 0.11 & $-0.10$ & 0.11 \\ [0.053cm]  
            38 & \ion{Sr}{ii} $^{\textsc{NLTE}}$            & 1   & 0.50  & 0.10 & 0.12 & $-0.11$ & 0.12 \\ [0.053cm]  
            38 & $\langle$ Sr $\rangle$ $^{\textsc{NLTE}}$  & 2   & 0.50  & 0.01 & 0.09 & $-0.11$ & 0.09 \\ [0.053cm]
            \hline
        \end{tabular}
            \hspace{0.5cm}
        \begin{tabular}{@{}c@{\hspace{9pt}}l@{\hspace{9pt}}c@{\hspace{9pt}}c@{\hspace{9pt}}c@{\hspace{9pt}}c@{\hspace{9pt}}c@{\hspace{9pt}}c@{}}
            \hline\hline
            \rule{0pt}{2.6ex} Z & Elem. & N & \logeps & $\sigma_{\rm red}$ & $\sigma_{\rm tot}$ & [X/Fe] & \sigXoFe \\ [0.1cm]
            \hline \\
            [-0.1cm]
            39 & \ion{Y}{ii} $^{\textsc{NLTE}}$  & 12  & $-0.29$     & 0.05 & 0.07 & $-0.19$     & 0.07 \\ [0.05cm]
            40 & \ion{Zr}{ii}                    & 22  & 0.43        & 0.02 & 0.07 & 0.16        & 0.07 \\ [0.05cm]
            41 & \ion{Nb}{ii}                    & 1   & $-0.52$     & 0.10 & 0.16 & 0.37        & 0.16 \\ [0.05cm]
            42 & \ion{Mo}{ii}                    & 1   & $-0.03$     & 0.10 & 0.17 & 0.36        & 0.17 \\ [0.05cm]
            44 & \ion{Ru}{i}                     & 3   & 0.01        & 0.06 & 0.12 & 0.48        & 0.12 \\ [0.05cm]
            45 & \ion{Rh}{i}                     & 1   & $-0.89$:    & 0.10 & 0.28 & 0.30:       & 0.28 \\ [0.05cm]
            46 & \ion{Pd}{i}                     & 1   & $-0.53$     & 0.10 & 0.11 & 0.12        & 0.11 \\ [0.05cm]
            47 & \ion{Ag}{i}                     & 1   & $-1.33$:    & 0.10 & 0.17 & 0.04:       & 0.17 \\ [0.05cm]
            49 & \ion{In}{ii}                    & 1   & $\leq$ 0.00 & /    & /    & $\leq$ 0.71 & /    \\ [0.05cm]
            56 & \ion{Ba}{ii} $^{\textsc{NLTE}}$ & 3   & $-0.23$     & 0.02 & 0.07 & $-0.09$     & 0.07 \\ [0.05cm]
            57 & \ion{La}{ii}                    & 9   & $-1.11$     & 0.02 & 0.05 & 0.07        & 0.05 \\ [0.05cm]
            58 & \ion{Ce}{ii}                    & 9   & $-0.63$     & 0.01 & 0.08 & $-0.02$     & 0.08 \\ [0.05cm]
            59 & \ion{Pr}{ii}                    & 6  & $-1.14$:    & 0.07 & 0.14 & 0.59:       & 0.14 \\ [0.05cm]
            60 & \ion{Nd}{ii}                    & 19  & $-0.60$     & 0.02 & 0.06 & 0.26        & 0.06 \\ [0.05cm]
            62 & \ion{Sm}{ii}                    & 9   & $-0.92$     & 0.01 & 0.04 & 0.39        & 0.04 \\ [0.05cm]
            63 & \ion{Eu}{ii} $^{\textsc{NLTE}}$ & 2   & $-1.21$     & 0.02 & 0.08 & 0.58        & 0.08 \\ [0.05cm]
            64 & \ion{Gd}{ii}                    & 26  & $-0.78$     & 0.03 & 0.12 & 0.42        & 0.12 \\ [0.05cm]
            65 & \ion{Tb}{ii}                    & 3   & $-1.43$     & 0.17 & 0.18 & 0.60        & 0.18 \\ [0.05cm]
            66 & \ion{Dy}{ii}                    & 19  & $-0.66$     & 0.01 & 0.05 & 0.51        & 0.05 \\ [0.05cm]
            67 & \ion{Ho}{ii}                    & 3   & $-1.56$     & 0.03 & 0.08 & 0.24        & 0.08 \\ [0.05cm]
            68 & \ion{Er}{ii}                    & 5   & $-0.77$     & 0.01 & 0.05 & 0.61        & 0.05 \\ [0.05cm]
            69 & \ion{Tm}{ii}                    & 3   & $-1.68$     & 0.04 & 0.15 & 0.63        & 0.15 \\ [0.05cm]
            70 & \ion{Yb}{ii}                    & 2   & $-1.00$:    & 0.00 & 0.09 & 0.23:       & 0.19 \\ [0.05cm]
            71 & \ion{Lu}{ii}                    & 1   & $-1.42$:    & 0.10 & 0.16 & 0.83:       & 0.16 \\ [0.05cm]
            72 & \ion{Hf}{ii}                    & 1   & $-1.14$     & 0.10 & 0.16 & 0.29        & 0.16 \\ [0.05cm]
            73 & \ion{Ta}{ii}                    & 3   & $\leq-0.75$ & /    & /    & $\leq$ 1.73 & /    \\ [0.05cm]
            76 & \ion{Os}{i}                     & 1   & $\leq-0.10$ & /    & /    & $\leq$ 0.96 & /    \\ [0.05cm]
            77 & \ion{Ir}{i}                     & 2   & -0.29:      & 0.01 & 0.13 & 0.64:       & 0.13 \\ [0.05cm]
            79 & \ion{Au}{ii}                    & 1   & $\leq$ 0.20 & /    & /    & $\leq$ 1.50 & /    \\ [0.05cm]
            82 & \ion{Pb}{i}                     & 2   & $\leq$ 0.30 & /    & /    & $\leq$ 0.61 & /    \\ [0.05cm]
            90 & \ion{Th}{ii}                    & 5   & $\leq-1.60$ & /    & /    & $\leq$ 0.65 & /    \\ [0.05cm]
            92 & \ion{U}{ii}                     & 4   & $\leq-1.00$ & /    & /    & $\leq$ 1.83 & /    \\ [0.05cm]
            \hline
        \end{tabular}
        \tablefoot{$Z$ is the atomic number, $N$ the number of lines used to derive the abundance, and $\log \epsilon$ the abundance. The reduced line-to-line scatter $\sigma_{\rm red}$ was computed as $\sigma_{\rm red} = \sigma_{\rm line}/\sqrt{N}$, where $\sigma_{\rm line}$ is the standard deviation from $N$ lines. When only one line was available, a conservative estimate $\sigma_{\rm red}=0.1$ dex was adopted. For elements measured from only two or three lines, the sample dispersion is poorly determined and may either overestimate or underestimate the true line-to-line scatter. Instead, we adopted a representative value, $\sigma_{\rm line, avg}=0.09$ dex, corresponding to the mean line-to-line dispersion derived from all elements measured with more than three lines. $\sigma_{\rm tot}$ is the total uncertainty, including the propagation of stellar parameters uncertainties on the  abundances. The [X/Fe] ratios and their associated uncertainties $\sigma_{\text{[X/Fe]}}$ are also reported. Abundances from different chemical species are reported separately, and $\langle {\rm X} \rangle$ denotes the mean abundance over all lines of a given element. Abundances derived using NLTE computations are flagged as “NLTE”. For some elements, only upper limits could be obtained, indicated by the symbol “$\leq$”. Carbon and nitrogen abundances were derived from molecular bands in the 4200-4400~\AA\ for C and around 3880~\AA\ for N. Values marked with “:” denote uncertain abundances, due to to weak or blended lines or to a non-optimal fit of the spectral region.}
    \end{table*}

\twocolumn

\newpage
\onecolumn

\section{Comparison with explosive nucleosynthesis yields}
    \begin{figure*}[h]
        \centering
        \includegraphics[width=\linewidth]{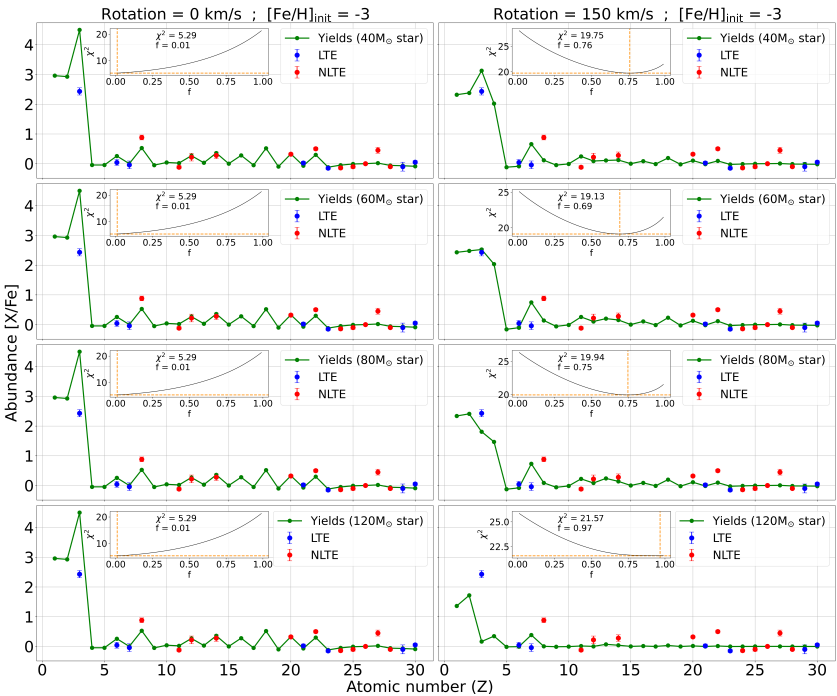}
        \caption{\tiny  Abundance pattern of \BHs\ (LTE in blue, NLTE in red) compared with explosive nucleosynthesis yields from \cite{Limongi2018} from core-collapse supernova (ccSN) with and without rotation velocities (0 and 150 km~s$^{-1}$), with initial masses of 40, 60, 80, and 120 \msun\ and an initial metallicity $\mathrm{[Fe/H]} = -3$ (green lines). In each panel, the inset shows the dilution factor $f$ (dilution into a material with a solar abundance pattern scaled to the metallicity of \BHs minimizing the reduced $\chi^2$). Lithium was excluded from the computation of the dilution factor.}
        \label{fig:AbProfile_vs_Yields}
    \end{figure*}

\onecolumn 
\section{Comparison of chemical abundances of \BHs\ with ED-2 stream members}\label{Sect: Appendix-ED2}

  Using archival ESO-UVES spectra, we derived abundances for four ED-2 stars using the same methodology adopted for \BHs. We adopted the stellar parameters from \cite{Balbinot2024}, which are in excellent agreement with isochrone fitting. Because of the limited S/N, only 4 to 6 elements could be measured in three stars, whereas 22 elements were determined for star 8896. Uncertainties were estimated from the line-to-line scatter (or fixed at 0.1 dex when $N_{\rm line}<4$), and temperature uncertainties were propagated using $\Delta$\teff\ values from \cite{Mucciarelli-2021}.

    \begin{table*}[h!]
        \caption{Comparison of chemical abundances of \BHs\ with four ED-2 stream members.}
        \centering
        \begin{tabular}{c@{\hspace{25pt}} c@{\hspace{25pt}}c@{\hspace{25pt}}c@{\hspace{25pt}}c@{\hspace{25pt}}c}
        \hline \hline
             & \BHs\ & 1904 & 4496 & 8896 & 5600 \\ [0.1cm]
             \hline \\ [-0.2cm]
            \teff [K] & $5255 \pm 65$ & $6657$ & $5974$ & $5620$ & $6110$ \\ [0.1cm]
            \logg & $2.93 \pm 0.21$ & $4.30$ & $4.53$ & $3.60$ & $4.55$ \\ [0.1cm]
            \vmic [km/s] & $1.39 \pm 0.07$ & $1.68$ & $1.12$ & $1.32$ & $1.28$ \\ [0.1cm]
            $\mathrm{[Fe/H]}$ & $-2.30 \pm 0.12$ & $-2.48 \pm 0.08$ & $-2.70 \pm 0.05$ & $-2.54 \pm 0.05$ & $-2.57 \pm 0.05$ \\ [0.1cm]
            \hline \\ [-0.2cm]           
            {[Na/Fe]} & $-0.12 \pm 0.0$5  &   /    &  /               & $0.02 \pm 0.03$  & / \\ [0.1cm]
            {[Mg/Fe]} & $0.22  \pm 0.13$  &   /    &  /               & $0.41 \pm 0.09$  & / \\ [0.1cm]
            {[Si/Fe]} & $0.28  \pm 0.11$  &   /    &  /               & $0.53 \pm  0.1$0 & / \\ [0.1cm]
            {[Ca/Fe]} & $0.32  \pm 0.04$  &   /    &  /               & $0.46 \pm  0.0$7 & / \\ [0.1cm]
            {[Ti/Fe]} & $0.50  \pm 0.05$  &   /    &  /               & $0.72 \pm  0.0$9 & / \\ [0.1cm]           
            {[Mn/Fe]} & $-0.10 \pm 0.06$  &   /    &  /               & $0.07 \pm 0.03$  & / \\ [0.1cm]    
            {[Ni/Fe]} & $-0.10 \pm 0.04$  &   /    &  /               & $0.07 \pm 0.13$  & / \\ [0.1cm]
            {[Zn/Fe]} & $0.05  \pm 0.05$  &   /    &  /               & $0.27 \pm 0.10$  & / \\ [0.1cm]
            {[Sr/Fe]} & $-0.11 \pm 0.09$  & $-0.09 \pm 0.11$ & $0.20  \pm 0.11$ & $0.09  \pm 0.12$ & $0.06  \pm 0.11$ \\ [0.1cm]
            {[Y/Fe]}  & $-0.19 \pm 0.07$  & $0.14  \pm 0.11$ & $-0.01 \pm 0.13$ & $-0.04 \pm 0.06$ & $-0.04 \pm 0.13$ \\ [0.1cm]
            {[Zr/Fe]} & $0.16  \pm 0.07$  & /                & $0.77  \pm 0.11$ & $0.45  \pm 0.06$ & $0.31  \pm 0.06$ \\ [0.1cm]
            {[Ba/Fe]} & $-0.09 \pm 0.07$  & $-0.06 \pm 0.11$ & $0.03  \pm 0.09$ & $0.11  \pm 0.06$ & $0.02  \pm 0.06$ \\ [0.1cm]
            {[La/Fe]} & $0.07  \pm 0.05$  & /                & /                & $0.37  \pm 0.06$ &  /               \\ [0.1cm]
            {[Nd/Fe]} & $0.26  \pm 0.06$  & /                & /                & $0.67  \pm 0.11$ &  /               \\ [0.1cm]
            {[Eu/Fe]} & $0.58  \pm 0.08$  & /                & /                & $0.79  \pm 0.06$ &  /               \\ [0.1cm]
            {[Tb/Fe]} & $0.60  \pm 0.18$  & /                & /                & $1.22  \pm 0.11$ &  /               \\ [0.1cm]
            {[Dy/Fe]} & $0.51  \pm 0.05$  & /                & /                & $0.84  \pm 0.06$ & $0.83  \pm 0.11$ \\ [0.1cm]
            {[Ho/Fe]} & $0.24  \pm 0.08$  & /                & /                & $0.64  \pm 0.08$ &  /               \\ [0.1cm]
            {[Er/Fe]} & $0.61  \pm 0.05$  & /                & /                & $0.95  \pm 0.08$ &  /               \\ [0.1cm]
            {[Yb/Fe]} & $0.23  \pm 0.09$  & /                & /                & $0.46  \pm 0.15$ & $0.66  \pm 0.11$ \\ [0.1cm]
            {[Lu/Fe]} & $0.83  \pm 0.16$  & /                & /                & $1.55  \pm 0.12$ &  /               \\ [0.1cm]
            {[Hf/Fe]} & $0.29  \pm 0.16$  & /                & /                & $0.64  \pm 0.11$ &  /               \\ [0.1cm]
            \hline
        \end{tabular}
    
        \tablefoot{The stars {\it Gaia} DR3 \texttt{source\_id} 4245\-5224\-6855\-409\-1904, 4479\-2263\-1075\-831\-4496, 6632\-3350\-6023\-108\-8896 and 6746\-1145\-8505\-626\-5600 are identified by their last four digits. Stellar parameters of the four ED-2 stream members are from \cite{Balbinot2024}.}
        \label{tab:placeholder}
    \end{table*}

\end{appendix}

\end{document}